\documentclass[sigconf]{acmart}

\usepackage{balance}
  
\usepackage{graphicx,url}
\usepackage[utf8]{inputenc}  
\usepackage{booktabs} 
\usepackage{graphicx}
\usepackage{environ}
\usepackage{tikz}
\usetikzlibrary{calc,matrix}
\usepackage{cleveref}
\usepackage{float}
\usepackage[normalem]{ulem}
\useunder{\uline}{\ul}{}
\usepackage{graphicx}
\usepackage{amssymb}
\usepackage{xcolor}
\usepackage{paralist}
\usepackage{gensymb}
\usepackage{balance}
\usepackage{float}
\usepackage[normalem]{ulem}
\useunder{\uline}{\ul}{}
\usepackage{graphicx}
\usepackage{amssymb}
\usepackage{xcolor}
\usepackage{paralist}
\usepackage{gensymb}
\usepackage{booktabs} 
\usepackage{subfigure}
\usepackage{booktabs}
\usepackage{multirow}
\usepackage{makecell}
\usepackage{url}
\usepackage{multirow}
\usepackage{graphicx}

\usepackage[show]{chato-notes}

\def\BibTeX{{\rm B\kern-.05em{\sc i\kern-.025em b}\kern-.08emT\kern-.1667em\lower.7ex\hbox{E}\kern-.125emX}}

\copyrightyear{2019} 
\acmYear{2019} 
\setcopyright{acmcopyright}
\acmConference[WebSci '19]{11th ACM Conference on Web Science}{June 30-July 3, 2019}{Boston, MA, USA}
\acmBooktitle{11th ACM Conference on Web Science (WebSci '19), June 30-July 3, 2019, Boston, MA, USA}
\acmPrice{15.00}

\acmSubmissionID{webs036}

\begin{document}

\title{Characterizing Attention Cascades in WhatsApp Groups}

\author{Josemar Alves Caetano}
\email{josemarcaetano@dcc.ufmg.br}
\orcid{0000-0002-0522-2572}
\affiliation{%
  \institution{Dept. of Computer Science, Universidade Federal de Minas Gerais (UFMG), Brazil}
}

\author{Gabriel Magno}
\email{magno@dcc.ufmg.br}
\affiliation{%
  \institution{Dept. of Computer Science, Universidade Federal de Minas Gerais (UFMG), Brazil}
}

\author{Marcos Gonçalves}
\email{mgoncalv@dcc.ufmg.br}
\affiliation{%
  \institution{Dept. of Computer Science, Universidade Federal de Minas Gerais (UFMG), Brazil}
}

\author{Jussara Almeida}
\email{jussara@dcc.ufmg.br}
\affiliation{%
  \institution{Dept. of Computer Science, Universidade Federal de Minas Gerais (UFMG), Brazil}
}

\author{Humberto T. Marques-Neto}
\email{humberto@pucminas.br}
\affiliation{%
  \institution{Dept. of Computer Science, Pontif\'{i}cia Universidade Cat\'{o}lica de Minas Gerais (PUC Minas), Brazil}
}

\author{Virg\'{i}lio Almeida}
\email{virgilio@dcc.ufmg.br}
\affiliation{%
  \institution{Dept. of Computer Science, Universidade Federal de Minas Gerais (UFMG), Brazil}}
\additionalaffiliation{%
\institution{Berkman Klein Center for Internet \& Society, Harvard University, USA}}

%
\renewcommand{\shortauthors}{J.A. Caetano et al.}

%
\begin{abstract}
An important political and social phenomena discussed in several countries, like India and Brazil, is the use of WhatsApp to spread false or misleading content. However, little is known about the information dissemination process in WhatsApp groups. Attention affects the dissemination of information in  WhatsApp groups, determining what topics or subjects are more attractive to participants of a group. In this paper, we characterize and analyze how attention propagates among the participants of a WhatsApp group. 
An attention cascade begins when a user asserts a topic in a message to the group, which could include written text, photos, or links to articles online. 
Others then propagate the information by responding to it. We analyzed attention cascades in more than 1.7 million messages posted in 120 groups over one year. Our analysis focused on the structural and temporal evolution of attention cascades as well as on the behavior of users that participate in them.  We found specific characteristics in cascades associated with groups that discuss political subjects and false information.  For instance, we observe that cascades with false information tend to be deeper, reach more users, and last longer in political groups than in non-political groups.

\end{abstract}


%

%

%
\maketitle

\section{Introduction}
Global mobile messenger apps, such as WhatsApp, WeChat, Signal, Telegram, and Facebook-Messenger, are popular all over the world. WhatsApp, in particular, with more than 1.5 billion users~\cite{15billion},  has an expressive penetration in many countries like South Africa, India, Brazil, and Great Britain. Messenger apps are an important medium by which individuals communicate, share news and information, access services and do business and politics. With end-to-end encryption added to every conversation, WhatsApp sets the bar for the privacy of digital communications worldwide. As a consequence, in many countries, WhatsApp has been used by political parties and religious activists to send messages and distribute news.  The combination of encrypted messages and group messaging has proved to be a valuable tool for mobilizing political support and disseminating political ideas~\cite{truckers2}.  There is growing evidence of unprecedented disinformation and fake news campaigns in WhatsApp. In India,  false rumors and fake news intended to inflame sectarian tensions have gone viral on WhatsApp, leading to lynching mobs and death of dozens of people~\cite{refintro1}.  Misleading messages with fake news and photos containing disinformation have been used to influence real-world behavior in the Brazilian elections in 2018~\cite{refresults1}. 

A key component of the WhatsApp architecture is group chats, which allow group messaging.  Groups are unstructured spaces where participants can share messages, photos, and videos with up to 256 people at once. Much of the political action in WhatsApp takes place in private groups and through direct messaging, which are impossible to analyze due to the encryption protocol. However, there is a large number of groups that are often publicized in well-known websites. These groups are open to new participants.  After joining a group, a participant can post messages and receive the messages that circulate in the group. Unstructured conversations are the most usual form of conversation in a group chat.

Differently, from Facebook and Twitter,  WhatsApp groups do not have many features that have been pointed out as responsible for the dissemination of false information~\cite{reffacebookvswhatsapp}. WhatsApp does not have ads, promoted tweets, News Feed or timeline, which are controlled by algorithms.  In a group chat, anybody can propose a discussion or start a new discussion in response to a piece of content or join an existing discussion at any time and any depth. There is no algorithm or human curator moderating discussions in a WhatsApp group. In such a domain, a natural question that arises is: What are the ingredients of the WhatsApp architecture that makes it a weaponizable social media platform to distribute misinformation?  The goal of this paper is to take a first step towards understanding information diffusion in WhatsApp groups. Our work is an essential step in order to increase knowledge about misinformation dissemination in messenger apps.

The approach we use to understand information spreading in a WhatsApp group is a combination of two concepts: information cascade and attention.  WhatsApp allows a  participant to explicitly mention a message she intends to reply or refer to, by using the reply function.  The reply function in WhatsApp creates a sequence of interrelated messages, which can be tracked and viewed as a cascade of interrelated messages. Every time a participant ``replies'' to a message, all other participants are notified by the message that appears on the screen.  A \emph{reply}  has the effect of re-gaining the attention for a  specific piece of information.  

In a chaotic environment of a WhatsApp group chat, participants are sometimes swamped by messages, which characterizes a typical information overload situation, that competes for the cognitive resources of the participants. Herbert Simon~\cite{Ciampaglia2015} was the one who first theorized about information overload, proposing the concept of economy of attention. ``A wealth of information creates a poverty of attention'' said  Simon~\cite{refintro3_Simon:Organizations}. \emph{Attention cascades} can be viewed as a structure that represents the process of information creation and consumption, which models new patterns of collective attention. They capture the connections fostered among group participants, their comments and their reactions in the group discussion setting.  We want to increase our understanding of information dissemination processes that underly the dynamics of a group discussion.  Our initial research questions are the following: 

\begin{itemize}
\item {\it How different are attention cascades in political and non-political groups on WhatsApp? 
\item  What is the impact of false information on the characteristics of attention cascades? }
\end{itemize}

To illuminate the behavior of attention cascades, we study the dissemination processes of cascades in political and non-political groups as well as cascades with previously reported false information and with unclassified information.  We show that the attention cascades are markedly different in their time evolution and structure.  More broadly, by characterizing the attention cascades of different types of groups, we can better understand the information spreading process in group discussions and explore the contextual factors driving their differences. 

In this paper, we present a large-scale study of attention cascades in WhatsApp groups.  Attention cascades can help to understand the process of information dissemination in WhatsApp groups.  With an in-depth analysis of a large number of messages of two groups, political and non-political, each containing verified false information and unclassified information,  we demonstrate that the structural and temporal characteristics of the two classes of cascades are different. We show that attention cascades in WhatsApp groups reflect real-world events that capture public attention. We also demonstrate that political cascades last longer than non-political ones.  We show that cascades with false information in political groups are deeper, broader and reach more users than the same type of cascades in non-political groups.

The remainder of this paper is organized as follows. We first review related work. Next, we discuss the concept of attention cascade that drives our study and describes each step of our methodology. We present our results for each research question posed, and then discuss our main conclusions and directions for future work.  

\section{Related work}
Characterizing cascades structure and growth on online social networks is a key task to understand their dynamics. Many studies have analyzed cascades formed on networks like Facebook and Twitter. In~\cite{Dow:2013}, authors analyzed large cascades of reshared photos on Facebook, more specifically one cascade of a reshared American President Barack Obama's photo posted on his page following his reelection victory, which in 2013 it had had over 4.4M likes, and another photo posted by a common Norwegian Facebook user (Petter Kverneng), which got one million likes on his humorous post. This paper shows that these two large cascades spread around the world in 24 hours and it points out that studying information cascades is important to understand how information is disseminated on today's very well-connected online environment. On the other hand, analyzing Twitter's cascades, the authors of~\cite{Rotabi:2017} analyzed the cascades from the point of view of users. This paper measures how Twitter users observe the retweets on their own timelines and how is their engagement with retweet cascades in counterpoint with non-retweeted content. Nevertheless, we need to understand cascades' properties on emerging online communication platforms like WhatsApp (which has millions of users organized in groups) to better understand the phenomenon of the quick information spread and its impact on users decisions and actions. 

Recently, the paper~\cite{Cheng:2018} shows how some diffusion protocols affect 98 large information cascades of photo, video, or link reshares on Facebook. Authors identify four classes of protocols characterized by the individual effort for user participates in a cascade and by the social cost of user stays out. The proposed classes, \textit{transient copy protocol}, \textit{persistent copy protocol}, \textit{nomination protocol}, and \textit{volunteer protocol}, may used to understand the cascade growth and structure. This paper indicates that as individual effort increase the information dissemination decreases and as higher the social costs users wait and observe more the group before making a decision or do any action.

The cascades can be viewed as complex dynamic objects. Thus, identifying their recurrence and predicting their growth are not trivial tasks. In~\cite{Cheng:2014:CP:2566486.2567997}, authors present a framework to address cascade prediction problems, which observe the nature of initial reshares as well as some characteristics of the post, such as caption, language, and content. Moreover, authors of~\cite{Cheng:2016:CR:2872427.2882993} analyze cascades bursts and show how their recurrence would occur over time. Predicting cascades growth, evolution, and recurrence could be useful to mitigate the spread of rumors~\cite{Friggeri:2014} and of misinformation in general, especially in the political context, which has received much attention lately.

Some studies analyzed attention on online platforms. Ciampaglia et al.~\cite{Ciampaglia2015} analyze Wikipedia network traffic to measure the attention span and their spikes through time. They found that collective attention is associated with the law of supply and demand for goods, in this case, information. Moreover, the authors show that the creation of new Wikipedia articles is associated with shifts of collective attention. Wu et al.~\cite{Wu17599} studies the collective attention towards news story from the political news webiste~\emph{digg.com}. The authors analyze how the group attention shifts considering the novelty of the news and the process of fade with time as they spread among people. 

Regarding WhatsApp, there are few studies analyzing data extracted from groups. The majority of works explore the WhatsApp effects on education activities. For instance, Cetinkaya~\cite{Cetinkaya2017fc} inspect the impact of WhatsApp usage on school performance and Bouhnik et al.~\cite{BouhnDeshe2014pg} analyze how students and teachers interact using WhatsApp. Other works compare WhatsApp with other applications. Church et al.~\cite{Church:2013:WUW:2493190.2493225} analyzes the difference of between WhatsApp and SMS messaging and in Rosler et al.~\cite{rosler2017more}, authors present a systematical analysis of security characteristics of WhatsApp and of two other messaging applications. These studies use qualitative methodologies and data related to (but not directly from) WhatsApp to perform the study. However, few works actually characterize WhatsApp usage using its data. Recently, Garimella et al. ~\cite{ICWSM1817865} proposes a data collection methodology and perform a statistical exploration to allow researchers to understand how public WhatsApp groups data can be collected and analyzed. Rosenfeld et al.~\cite{Rosenfeld_39_22} analyze 6 million
encrypted messages from over 100 users to build demographic prediction models that use activity data but not the content of these messages. In~\cite{moreno2017whatsapp}, authors collected WhatsApp messages to monitor critical events during the Ghana 2016 presidential elections. 

This paper greatly builds on top of prior study~\cite{Vosoughi1146}, where the authors analyzed the diffusion of verified true and false news stories distributed on Twitter. They classified news as true or false using information from six independent fact-checking organizations. They found that falsehood diffused farther, deeper, and more broadly than the truth in all categories of information, especially for false political news. They also found that false news was more novel than true news and false stories inspired fear, disgust, and surprise in replies, whereas true stories inspired anticipation, sadness, joy, and trust.

In our work, we also adjust the methodology proposed in~\cite{Vosoughi1146} on WhatsApp to analyze cascades properties and user relationships of political and non-political groups. We also analyze and compare cascades with and without identifiable falsehood using the information on Brazilian fact-checking websites. Nevertheless, we extend the work~\cite{Vosoughi1146} by applying the methodology on WhatsApp groups that are entirely different online environments when we compare to Twitter or Facebook. Finally, we exploit an \textit{attention perspective} in our analyses, which brings a somewhat different view of the problem of misinformation dissemination.

\section{Attention Cascades in WhatsApp} \label{sec:attentioncascade}
WhatsApp enables one-to-one communication through private chats as well as one-to-many and many-to-many communication through groups, allowing users to send textual and media (image, video, audio)  messages.  Compared to other online social networking applications (e.g., Facebook, Twitter), WhatsApp may be considered an unsophisticated platform as information is shared through a very loosely structured interface. Shared content is shown in temporal order, with each message accompanied by the posting time and, in case of groups, its origin (identification or telephone of the user who posted it). Yet, WhatsApp is a major communication platform in various countries of the world, offering an increasingly popular vehicle for information dissemination and social mobilization during important events~\cite{truckers2}.

One important feature of WhatsApp is \emph{reply}, which allows a user to explicitly mention a  message she intends to reply or refer to, bringing it forward in a conversation thread. This kind of interaction among users mentioning, replying or sharing each other's messages is common in other online social networks (e.g., reply and retweet in Twitter, share and post on Facebook). However, 
in WhatsApp this feature plays a key role in facilitating navigation through the sequence of messages, allowing one to keep track of different ongoing conversation threads. This is particularly important in the case of WhatsApp groups, where different subsets of the participating users may engage in different, possibly weakly related (or even unrelated) conversations simultaneously. Thus it becomes hard to keep track of each conversation thread, losing attention and ultimately diverting.   

Our goal in this paper is to study the information dissemination process in WhatsApp groups, focusing mainly on how user attention, a key channel for such process, is characterized in such unstructured, yet increasingly popular environment. To that end, we use the concept of \emph{ cascades}~\cite{Vosoughi1146} as a structural representation of how users' attention is dedicated to different conversation threads within a WhatsApp group.  We emphasize that conversations in WhatsApp have the distinguishing characteristic of \emph{ not} being influenced or driven by any algorithm as in other social networks
(e.g.,  newsfeed in the Facebook, content recommendation, etc.). They depend solely on the active participation of users of the group, as messages posted by others catch their attention. 

More specifically, an \emph{attention cascade} begins when a user makes an assertion about a topic in a message to the group. This message is the root of the cascade. Other users join and establish a conversation thread by explicitly replying to the root message or to other messages that replied to it. We say that the root or subsequent messages in the cascade \emph{caught the attention} of a group member, motivating her to interact. Thus, we focus on messages that were explicitly linked by the \emph{reply} feature\footnote{Users may not necessarily use the reply feature to respond to a previous message. However identifying such implicit links would require processing the content of the messages, which is outside the present scope.}. We consider the specific use of the reply feature as a  signal that the user's attention was caught by the message she is replying to and that the cascade is a (semi-)structured representation modeling emergent patterns of {\it collective} attention.  As in any real conversation of a group of people,  attention may drift to other (possibly weakly related) topics as the conversation goes on.  Note however that in an explicit reply, the message that is being replied to is shown to everyone, just before the reply itself, serving as a means to ``re-gain'' or keep (part of) the original attention that motivated the user to participate in the cascade. Thus, the root message triggers the initial attention of some users, and the conversation is kept alive among a subset of participating users (which may change over time)  as they continue replying to successive messages.  



We also note that, unlike prior uses of the cascade model to understand the propagation of a particular piece of content~\cite{Rotabi:2017,Dow:2013}, we here use it to represent the participation (thus attention) of multiple users in an explicitly defined conversation thread, despite other messages and conversations that might be simultaneously happening in the same group.    Thus, we refer to such structural representation of how collective attention is dedicated to a particular conversation thread within a WhatsApp group as \emph{attention cascade}.

\section{Methodology} \label{sec:methodology}
In this section, we describe our methodology to gather data and characterize attention cascades in WhatsApp groups. We analyze cascades concerning three dimensions, namely, structural and temporal properties as well as user participation. The first two capture the main patterns describing how collective attention is dedicated to a conversation thread and how it evolves. The latter captures how individuals participate in such process, interacting with each other via replies.    Given the extensive use of WhatsApp for discussions related to politics and other social movements~\cite{truckers2}, we collected data from groups oriented towards political and non-political topics. We define the topic of a cascade as the topic of the group it belongs to, and we analyze political and non-political cascades separately. Similarly, we also analyze cascades with identified false information separately from the rest.

In the following, we start by describing our methodology to collect WhatsApp data (Section \ref{subsec:collection}), and present how we identify and build cascades from such data (Section \ref{subsec:identifying}).  We discuss our group and cascade labelling method (Section \ref{subsec:labelling}) and present the attributes used in our characterization (Section \ref{subsec:attributes}).

 \subsection{Data Collection} \label{subsec:collection}
We collected a WhatsApp dataset consisting of all messages posted in  120 Brazilian WhatsApp groups from October $16^{th}$ 2017 to November $6^{th}$ 2018. Despite offering private chats by default, WhatsApp allows group administrators to share invites to join such groups on blogs, web pages, and other online platforms. Such feature effectively turns the access to the group and the content shared in it public since anyone who has the invite can choose to join the group, receiving all the messages posted after that. 

We focus on groups whose access is publicly available. To find invites to such groups online, we leveraged the fact that all links shared by group administrators have the term \texttt{chat.whatsapp.com} as part of their URLs. Thus, we used Selenium scripts on Google Search website to search for web pages containing this term and located in Brazil (according to Google). We parsed all pages returned as a result of the search, extracting the URLs to WhatsApp groups. 

The monitoring and data collection process of each group requires an actual device and valid SIM card to join the group. Thus we were restricted to join a limited number of WhatsApp groups by the available memory in our devices. 
We randomly selected and joined 120 groups to monitor. These groups cover different topics which can be further categorized into political oriented subjects or not, as will be discussed in Section \ref{subsec:labelling}.
We then used the same process described in~\cite{ICWSM1817865} to collect data from those groups,  preserving user anonymity and complying with WhatsApp privacy policies. Specifically, we extracted the WhatsApp messages from our monitoring devices (smartphones) using scripts provided by the authors\footnote{https://github.com/gvrkiran/whatsapp-public-groups}, replacing each unique telephone number by a random unique ID. Throughout the paper, we refer to such a unique ID by the term  \emph{user}\footnote{We are not able to identify multiple telephone numbers belonging to the same person.}. 

In total,  our dataset consists of 1,751,054 messages posted by 30,760 users engaged in the 120 WhatsApp as mentioned earlier groups~\footnote{Readers can contact the first author to request the anonymized dataset used in this work. We would be pleased to provide it.}. 

\subsection{Attention Cascade Identification} \label{subsec:identifying}
Recall that an attention cascade is composed by a tree of messages, which are pairwise connected by the reply feature, posted by one or more users.  
We identify cascades on WhatsApp by taking the following approach.  We model each group as a directed acyclic graph where each node is a message (text or media content) posted in the group and there is a direct edge from message \(A\) to message \(B\)  if message \(B\) is a reply to message \(A\)\footnote{Thus, message \(B\) was necessarily posted after message \(A\).}. 
 Each connected component of such a graph, which by definition is a tree,  is identified as a cascade. 
 Note that we chose not to impose any time constraint on the cascade identification because we consider the use of a reply as strong evidence of the connection between messages. Thus, all cascade leaves are messages that did {\it not} receive any reply (in our dataset).   
 
 \begin{figure}[ht]
    \centering
    \includegraphics[width=0.47\textwidth]{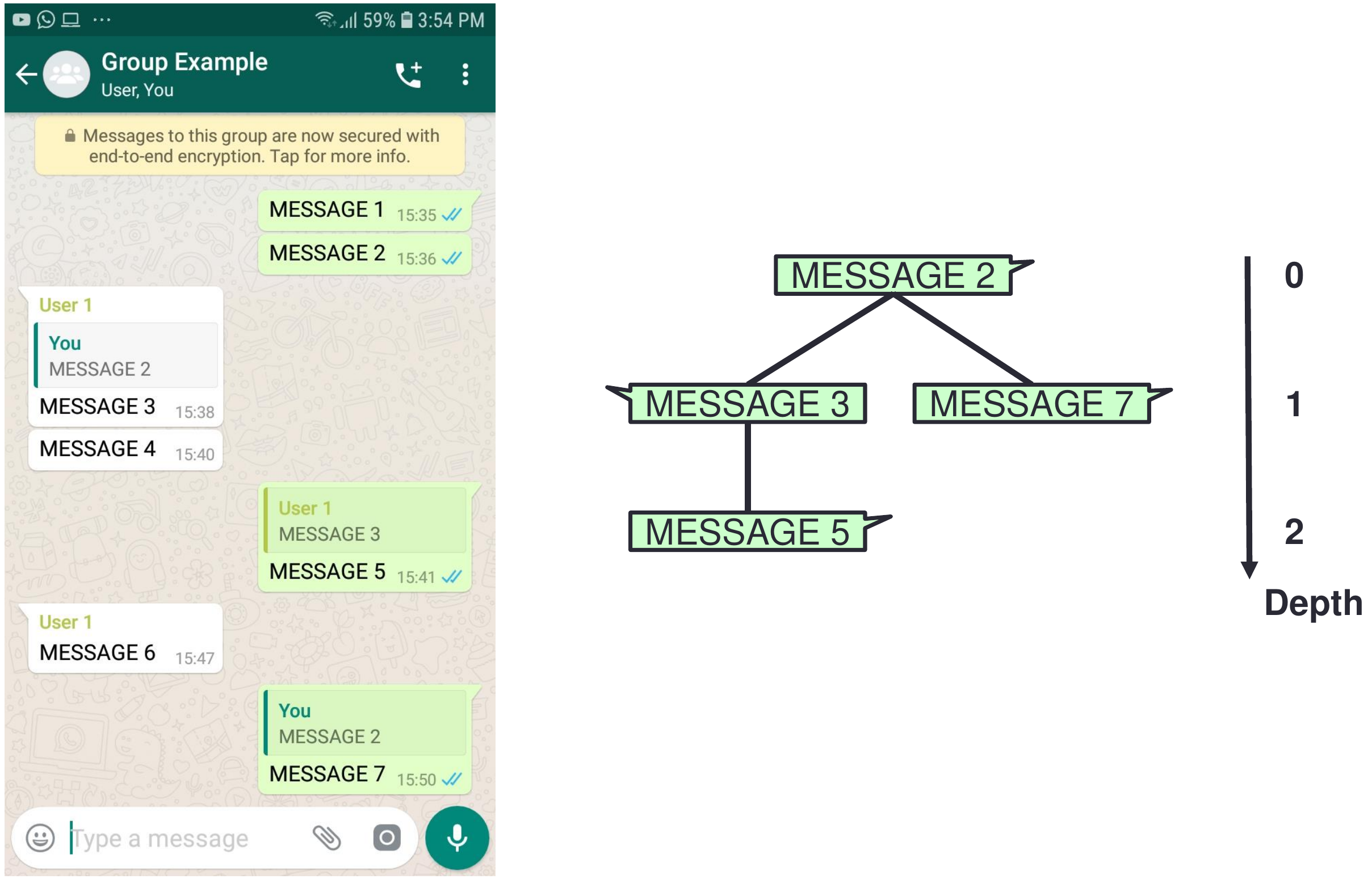}
\caption{Example of a cascade on WhatsApp.} 
\label{fig:cascades_sample}
\end{figure}
 
Figure~\ref{fig:cascades_sample} illustrates how we identify cascades from messages posted in a WhatsApp group.  Figure~\ref{fig:cascades_sample} (left) shows a sequence of 7 messages. The cascade starts with   ``MESSAGE 2'',   the root, which was posted at time 15:35. At time 15:38, another user replied to that message by posting ``MESSAGE 3'', which was then followed by another reply, ``MESSAGE 5''. At time 15:50,  user ``You'' replied the root message by posting ``MESSAGE 7''.  Note that  ``MESSAGE 4'' and ``MESSAGE 6'' posted during this time interval do not belong to the cascade as they are not explicit replies to any previous message in the cascade. Moreover, ``MESSAGE 1'' does not belong to the cascade either, since the message was posted before the root message. Moreover, since it did not receive any reply, it does not belong to {\it any} cascade. Finally, note that ``MESSAGE 7'' is a reply made by the same user who posted ``MESSAGE 2'', thus it is a self-reply. By such an example, it becomes clear that there might be multiple on-going cascades simultaneously in the same group.

\subsection{Classifying  Groups and Cascades}\label{subsec:labelling}

As mentioned, we analyze cascades of different topics, notably political and non-political topics,  as well as cascades containing false information separately, aiming at identifying attributes that distinguish them. In this section, we discuss how we label each cascade according to such categorization.

 WhatsApp groups are identified by names that reflect the general topic that the discussions in the group should cover. Examples of group names in our dataset (translated to English) are \emph{ Brazilian news}, \emph{ Friendship without borders} and \emph{ Right wing vs. Left wing}. Our data collection span several periods of significant social mobilization in Brazil. For instance, the 2018 Brazilian general election,  during which WhatsApp was reportedly a significant vehicle of political debates\footnote{Another example was a national truck drivers strike which greatly affected the country with the shutdown of highways in most states which caused unavailability of food and medicine for two weeks in late May 2018.}. Thus, we chose to classify groups into two categories: political oriented and non-political. To that end, we manually labeled the groups according to their names. 
 For instance, group names that refer to specific candidates or political debates were labeled as political groups, whereas all others were labeled as non-political. Out of the three examples above, the first two illustrate non-political groups, whereas the last one
 (\emph{Right wing vs. Left wing}) is clearly politically oriented. Our labeling effort identified  78 political and 42 non-political groups in our dataset. 
 
 We refer to an attention cascade in a political group as a \emph{ political cascade}. Similarly, cascades in non-political groups are labeled as  \emph{ non-political}. Such definition is a simple cascade categorization since the specific topic(s) discussed in a cascade may diverge from the group's meant subject. Nevertheless, we argue that, in general, inheriting the group label is a reasonable approximation, and we leave for the future a more specific content-based cascade categorization. 

Orthogonally, we also distinguish cascades containing reportedly false information (e.g., rumor) or, more generally speaking, {\it falsehood}, from the others. To that end, we relied on previously identified fake news reported by six Brazilian fact-checking websites (\textit{Aos Fatos}, \textit{Agência Lupa}, \textit{Comprova}, \textit{Boatos.org}, \textit{Checagem Truco}, and \textit{Fato ou Fake})~\footnote{\url{https://aosfatos.org/}, \url{https://piaui.folha.uol.com.br/lupa/}, \url{https://projetocomprova.com.br/},\url{https://www.boatos.org/},\url{https://apublica.org/checagem/},\url{https://g1.globo.com/fato-ou-fake/}}.  We first collected all news identified as fake by those six Brazilian fact-checking websites, totalling 3,072 fake news fact-checks. Next, we filtered fake news in text, that is, we removed fact-checks that analyzed fake content only in media formats (audio, video or image). In total, we selected 862 fake news texts.  

We then turned to the messages in our identified cascades, focusing on textual content and URLs shared. The latter often refers to news webpages. Thus, we extracted the texts from those URLs using the \texttt{newspaper}\footnote{\url{https://pypi.org/project/newspaper/}} Python library, which is specialized to retrieve the textual content of news portals and websites. We discarded all URLs for which the library was not able to recover the text (returned {\tt NULL}). 

In total, we identified 337,861 pieces of text on the cascades (327,530 textual messages and 10,331 texts from URLs). For the sake of readability, we refer to all of them as merely text messages.

In order to identify text messages with falsehood, we first processed all fake news and text messages by removing stopwords and performing lemmatization using the Natural Language Toolkit~\cite{bird2009natural}.  We then modeled each piece of content (fake news and text message) as a vector of size  \(n\), where \(n\) is the number of distinct lemmas identified in all text messages and fake news. Each position in the vector contains the weight of the corresponding lemma, defined as the term frequency (i.e., the number of times that the term appeared in the corresponding text).

We compared each text message against each fake news by computing the cosine similarity~\cite{Huang08similaritymeasures} between their vector representations.  Specifically, given a text message (vector) $m$ and fake news (vector) $f$, we computed their textual similarity as:

\begin{equation}\label{eq:cosine_similarity}
similarity(m,f) = \frac{m \cdot f}{\left \| m \right \|\left \| f \right \|} = \frac{\sum_{i=1}^{n} m_{i} f_{i}}{\sqrt{\sum_{i=1}^{n} m_{i}^2} \sqrt{\sum_{i=1}^{n} f_{i}^2}}
\end{equation}

\noindent where \(m_{i}\) and \(f_{i}\) are the weights  in text vectors \(m\) and \(f\). The cosine similarity returns a value between 0 and 1. The closer the value to 1, the stronger the similarity between the vectors.

In order to reduce noise, we focused on text pairs with similarity above 0.5. We manually inspected each such text pair and identified a total of  677 out of 2271 messages whose textual content referred to fake news reported by Brazilian fact-checking websites. 
We label any cascade containing at least one of such messages as {\it cascade with falsehood}\footnote{Although we do not impose restrictions on the depth of such message, we note that in the vast majority of the cases (93\%), the root message of such cascades carried false information.}.  We refer to all other cascades as {\it unclassified}. Note that we cannot guarantee the absence of falsehood in the unclassified cascades as we only analyzed fake news in text format, and we are restricted by the available fake news in such format reported by fact-checking websites. Nevertheless, we assume  that  the most popular   fake fake news  disseminated as textual content were identified and listed by Brazilian fact-checking services.

In total, we identified 666 cascades with falsehood (49 non-political and 617 political cascades) and 149,528 unclassified ones (52,689 non-political and  96,839 political cascades).

\begin{figure}[ht]
    \centering
    \includegraphics[width=0.47\textwidth]{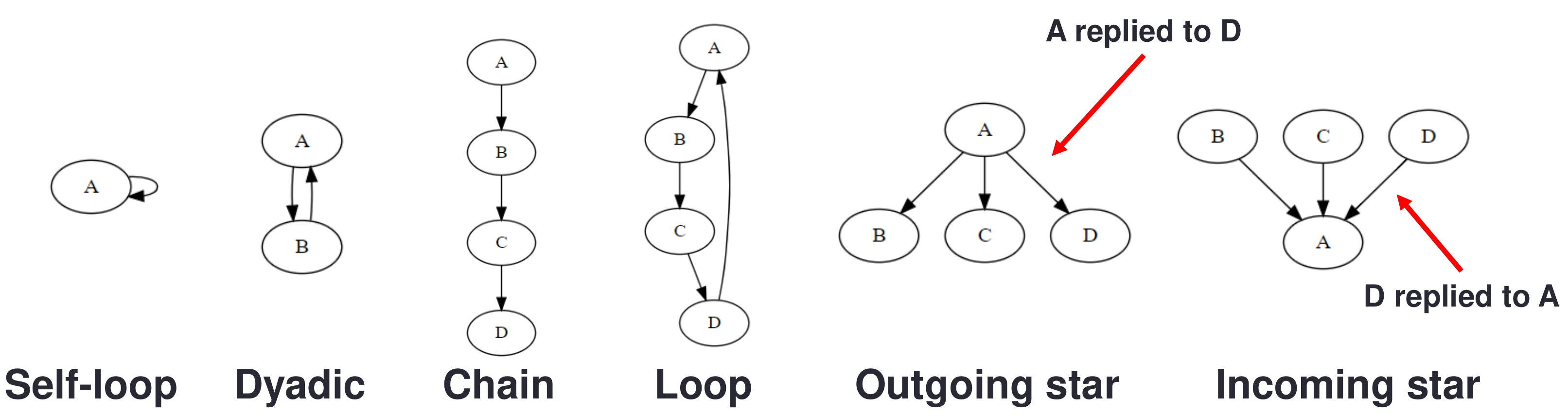}
\caption{Six motifs considered for analysis. Each node is a user and an edge ($i,j$) represents that a user $i$ replied a user $j$ in a given cascade.  }
    \label{fig:user_motifs_sample}
\end{figure}

\subsection{Cascade Attributes}\label{subsec:attributes}

We characterize attention cascades concerning several attributes which can be grouped into three main dimensions: structural properties, temporal properties, and user participation. The structural properties consist of the number of nodes, depth, maximum breadth, and structural virality. The number of nodes corresponds to the total number of messages in the cascade.  The depth of a particular message $m$ in a cascade is the number of edges from the root message to $m$ (i.e., number of replies from $m$ back to the root). The depth of a cascade is then defined as the maximum depth reached by a message in the cascade. The breadth of a cascade is defined according to a certain depth and corresponds to the number of messages in that particular depth of the cascade. Thus, the maximum breadth of a cascade is defined based on the breadth at all depth levels in the cascade.
The structural virality of a cascade is defined as the average distance between all pairs of nodes \cite{doi:10.1287/mnsc.2015.2158}. The higher the structural virality value, the stronger the indication that, on average, the messages are distant of each other, thus suggesting viral diffusion of content in that cascade~\cite{doi:10.1287/mnsc.2015.2158}.  

The main temporal attribute analyzed is the cascade duration, defined as the time interval between the last message belonging to the cascade and its root. We also characterize the temporal evolution of each cascade by analyzing its structural properties over time. When analyzing structural and temporal attributes that are in function of another attribute, we normalize their values in order to compare cascades with different values ranges. For example, when we analyze depth over time, we averaged (and measured standard error) at every depth considering the maximum depth of the cascade. Thus 50\% depth means the depth at the middle of any cascade. Therefore, we can analyze the average number of minutes it took to reach a certain percentage depth in a cascade and compare with others.

\begin{figure*}[ht]
    \centering
    \includegraphics[width=0.9\textwidth]{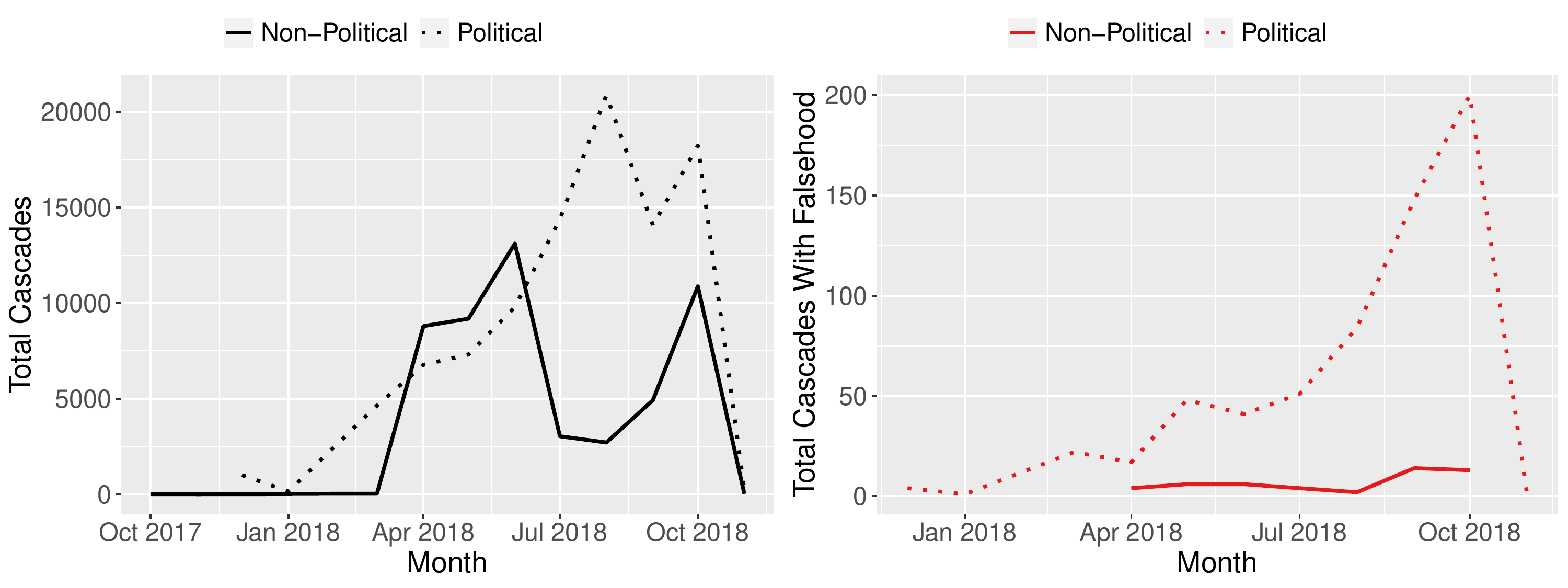}
\caption{Total cascades over time.}
    \label{fig:cascades_temporal_totals}
\end{figure*}

Regarding the third dimension,  user participation, we characterize each cascade in terms of the number of unique users participating in the cascade as well as the patterns of social communication established by them. For the latter, we first build a user-level representation of each cascade, that is,  we build a directed graph whose   vertices are users and  an edge between vertice \(X\) and vertice \(Y\) is added if user \(X\) replied to user \(Y\) {\it in the given cascade}.  We identify patterns of social communication occurring between users in such representation using motifs. To that end, we borrow from~\cite{Zhao:2010:CMT:1871437.1871694}, where the authors proposed network motifs as a tool to analyze patterns of information propagation in social networks. The authors proposed four types of motifs (i.e., long chain, ping pong, loop, and star), we here extend the definition by analyzing six motifs, which are shown in Figure~\ref{fig:user_motifs_sample}. Specifically, we propose the self-loop motif and divide the star motif into two other motifs by differentiating it according to its edge orientation (incoming star and outgoing star). Moreover, we choose to rename long chain to chain and ping pong to dyadic.  

We identify motifs in cascades testing if a cascade graph is isomorphic to a motif template. We define  templates for the motifs in Figure \ref{fig:user_motifs_sample} as follows. Self-loop and dyadic motifs are considered templates, as the total number of nodes is fixed (one and two, respectively). For the remaining motifs, we generate templates with the number of nodes ranging from 2 (chain) or 3 (loop, outgoing star, incoming star) up to \(n\), where \(n\) is the maximum number of unique users in any cascade analyzed. To detect graph isomorphism,  we used the implementation of the  VF2 algorithm proposed by Cordella {\it et al.}~\cite{1323804} available in the \emph{networkx} Python library. The VF2 algorithm identifies both graph and subgraph isomorphism. Our analysis considers the presence of a motif in a given cascade if the cascade is a graph (or subgraph) isomorphic to the corresponding template.


\section{Cascade Characterization}
In order to understand the characteristics of the attention cascades as well as the hidden structures existing in the interaction between participants of a group, we adopt a three-dimension approach to characterize attention cascades. To that end, we look at the structural and temporal characteristics of the cascades and the different communication forms that participants of a cascade interact with each other. In the following, we present the analysis and discussion of the findings in each dimension.

\begin{figure*}[ht]
    \centering
    \includegraphics[width=\textwidth]{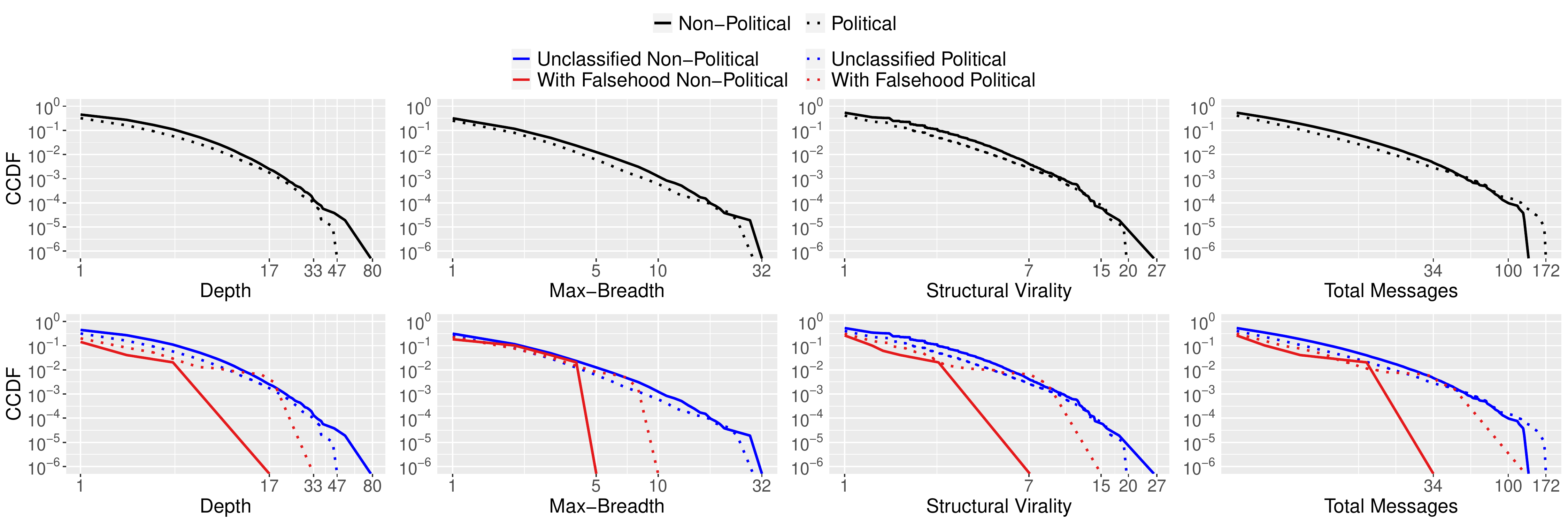}
\caption{Depth, maximum breadth, structural virality, and total messages.
}
    \label{fig:cascades_static_attributes_part_all}
\end{figure*}

\begin{figure*}[ht]
    \centering
    \includegraphics[width=0.9\textwidth]{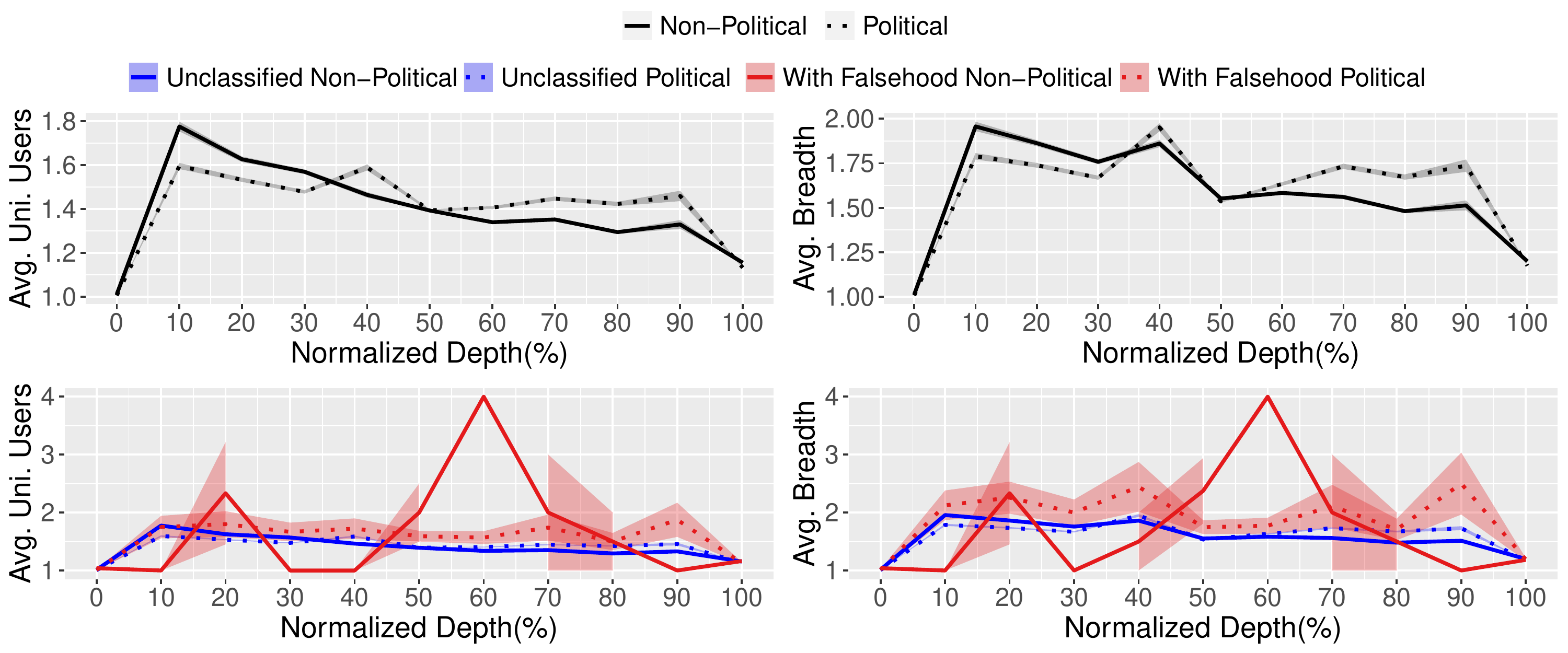}
\caption{Breadth vs. depth and unique users vs. depth of cascades from political and non-political WhatsApp groups.
}
    \label{fig:cascades_dynamic_attributes_structural}
\end{figure*}

\subsection{Overall cascade analysis}

In this section, we look at the number of cascades identified in our dataset of messages. To do so, and rather than viewing the set of all cascades, we segment the cascades into two classes, namely: political and non-political. For each class, we separate the cascades into two sub-classes: cascades with verified false information and unclassified cascades, that may have both true and false information, but it was not verified. 

\begin{figure*}[ht]
    \centering
    \includegraphics[width=0.7\textwidth]{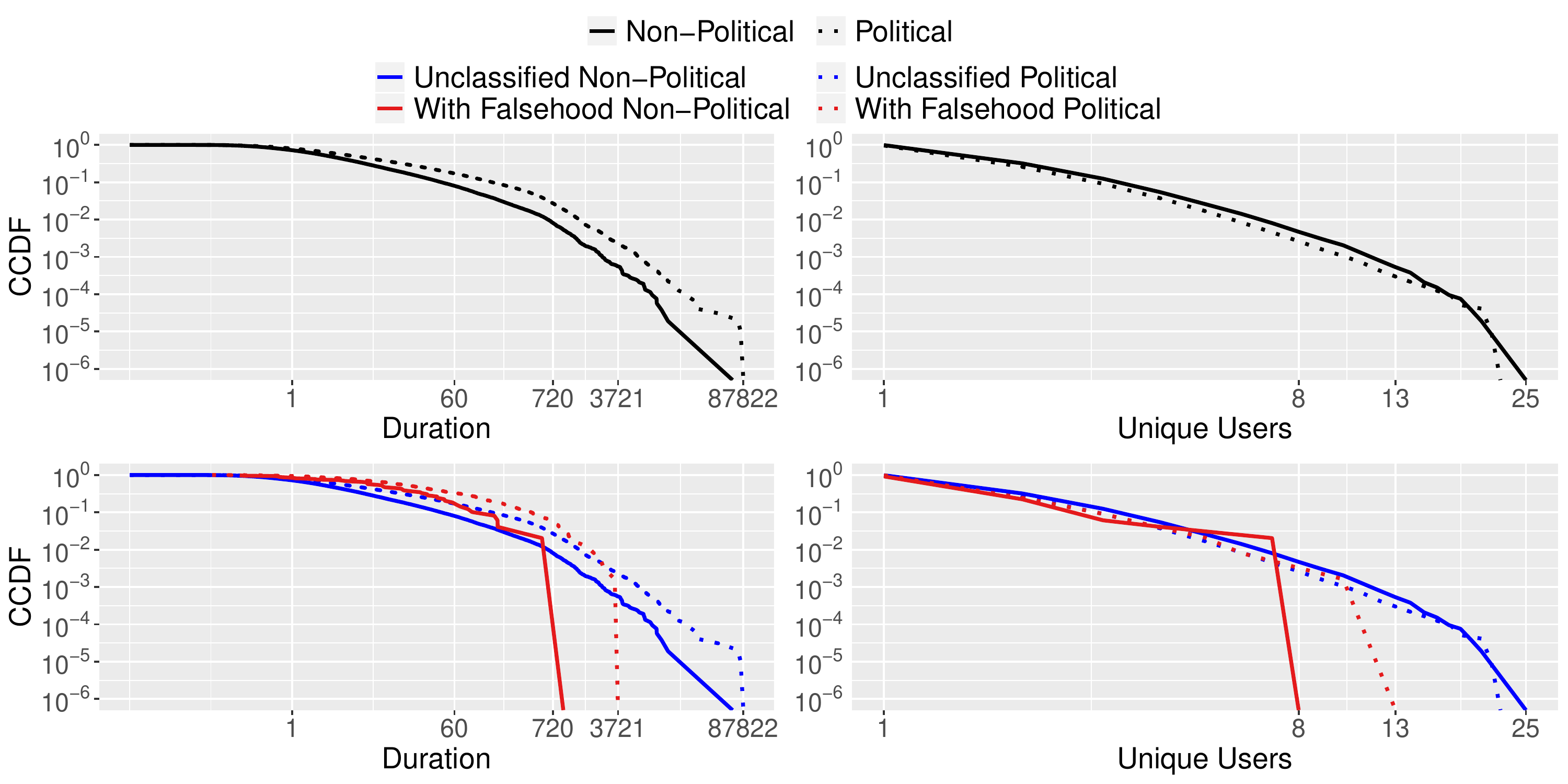}
\caption{Cascade duration and unique users. 
}
    \label{fig:cascades_static_attributes_part5}
\end{figure*}


\begin{figure*}[ht]
    \centering
    \includegraphics[width=0.9\textwidth]{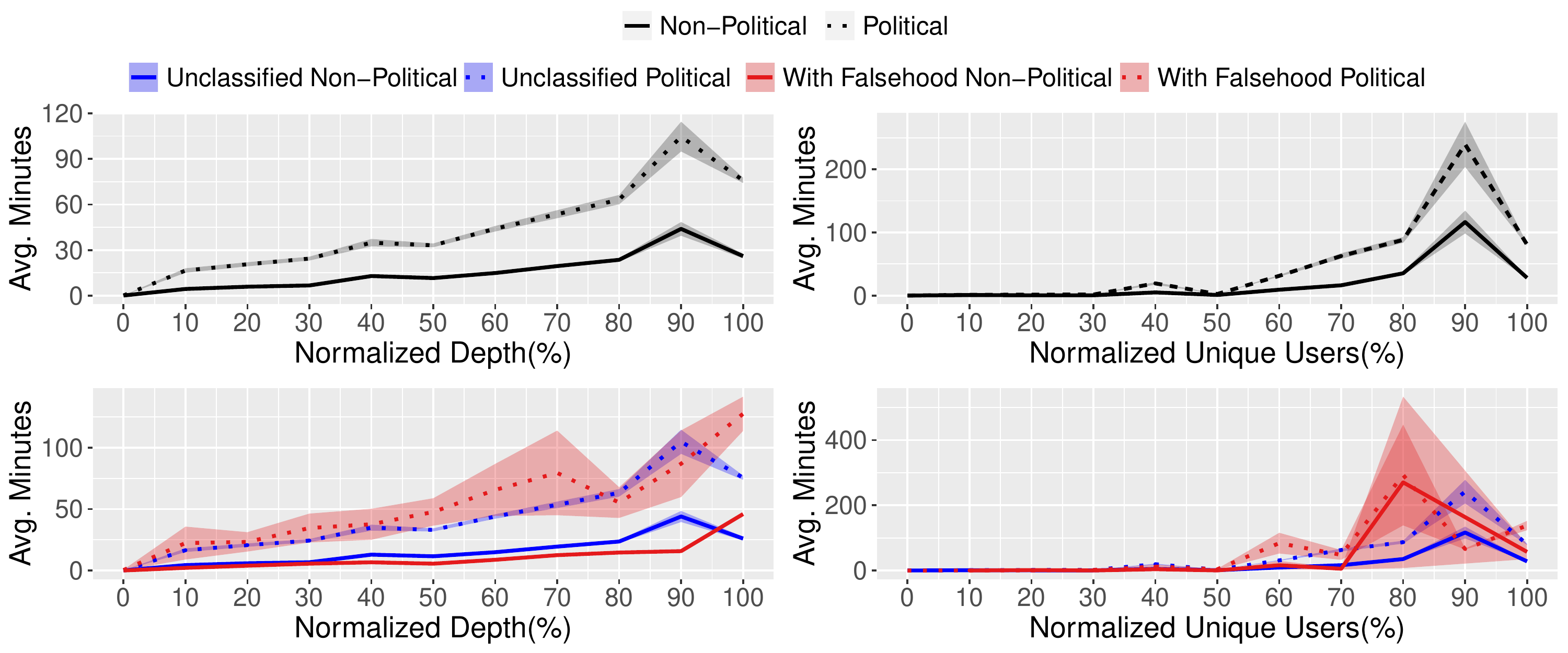}
\caption{Depth over time and unique users over time of cascades from political and non-political WhatsApp groups. 
}
    \label{fig:cascades_dynamic_attributes_time}
\end{figure*}

\begin{figure*}[ht]
    \centering
    \includegraphics[width=\textwidth]{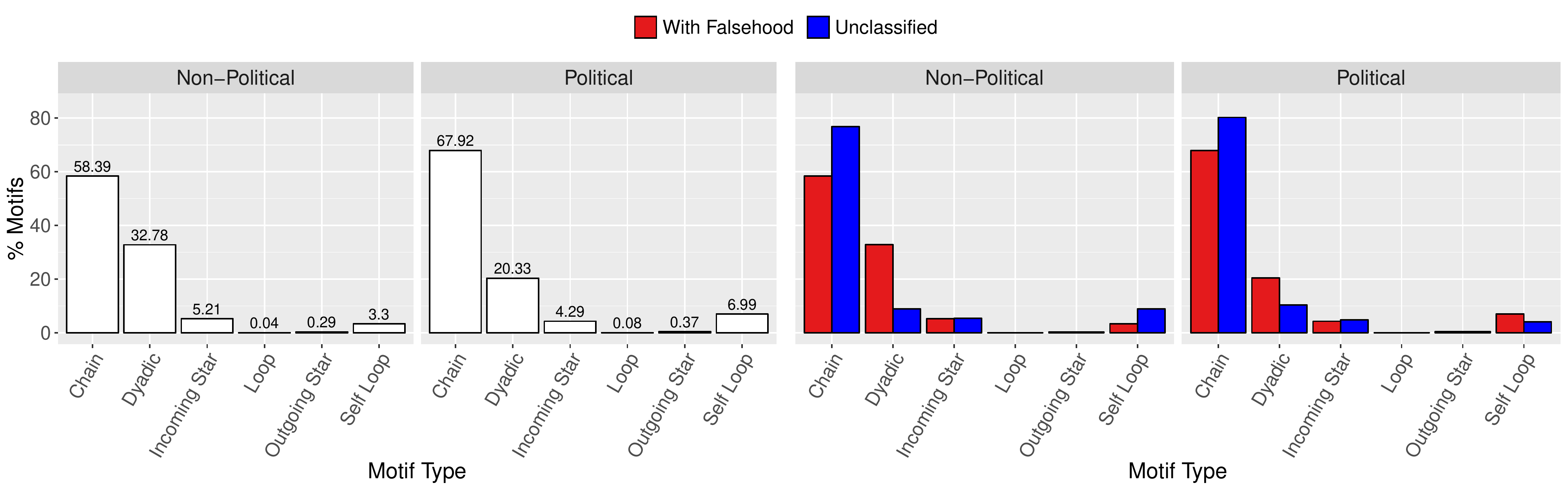}
\caption{User motifs.}
    \label{fig:user_motifs}
\end{figure*}


Figure~\ref{fig:cascades_temporal_totals} shows the total number of cascades in the dataset collected from WhatsApp groups during the period from October  2017 to November 2018. 
As it is evident from Figure~\ref{fig:cascades_temporal_totals}, there are two peaks of cascades in political groups. These peaks shown in the leftmost graph are associated with key events of the electoral campaign in Brazil. One peak occurred when the presidential candidate,  Jair Bolsonaro, was stabbed in the stomach during a campaign rally at the beginning of September. The other event was the day of the first round of voting, at the beginning of October. It is worth noting that the peak of the number of political cascades with false information is also associated with the date of the first round of voting, that was widely publicized by the media~\cite{refresults1}. The non-political cascades with falsehood remained stable during the election period, as exhibited in the rightmost graph of Figure~\ref{fig:cascades_temporal_totals}.  In the leftmost graph, the peak for non-political cascades occurred by the end of May, coinciding with the truckers’ strike that halted Brazil,  with protesters blocking traffic on hundreds of highways~\cite{refresults2}. These two graphs clearly show that attention cascades in WhatsApp groups reflect real-world events that capture the public attention.

We also analyzed temporal relationships among different cascades of each group~\cite{Allen:1983:MKT:182.358434}. We found that in 99.2\% of the cases there is no temporal overlap (i.e., one cascade finishes before others start). Therefore, our dataset shows that most of the time only one thread of discussion dominates the attention of the group.

\subsection{Structural Characteristics}

In this section, we focus on the structural characteristics of the cascades, that are defined in section~\ref{subsec:attributes}. Figure~\ref{fig:cascades_static_attributes_part_all} shows the  CCDF of the empirical distribution of the main structural characteristics of the two classes of cascades, political and non-political and the sub-classes, represented by cascades with false information and unclassified cascades. The upper row shows the comparison between political (dotted) and non-political (regular) groups. The lower row shows the comparison between `unclassified' (blue) and `with falsehood' (red), besides the previous segmentation between political and non-political, totaling four lines of distribution.

A natural question concerning the structure of attention cascades in different WhatsApp groups is whether they exhibit a similar structural profile. When analyzing the cascades on political and non-political groups (upper row) in Figure~\ref{fig:cascades_static_attributes_part_all}, we observe that the graphs show similar curves for both classes, with small differences in the probability of the maximum values for the two classes. We observe that non-political cascades are deeper and broader than political cascades.  Moreover, despite the structural virality of political and non-political cascades are very similar, when we look at the graph that shows the structural virality of the sub-classes of cascades (lower row), we note that the virality of cascades with false information in political groups is more significant than the virality in a non-political group. Recall that higher virality suggests viral diffusion, which maybe interpreted as a sign that false information in political groups spread farther than in non-political groups. Finally, the total number of messages attribute calls our attention since the number of messages in political discussions exceeds the number in non-political discussions, indicating that political subjects stimulate more interaction among participants.

Now we analyze the cascade attributes as a function of depth. We observe in Figure~\ref{fig:cascades_dynamic_attributes_structural} that cascades of political groups have on average more users than cascades of non-political groups after 35\% of the maximum depth. 
The cascades in political groups are on average wider than non-political ones after the half of the maximum depth, indicating that cascades of political groups finish with more parallel interactions (branches) than the non-political ones. Regarding misinformation comparison, we note that cascades with falsehood in political groups reach on average more users than the other types of cascades for higher depth. Note that cascades with falsehood of non-political groups have an unstable behavior. The cascades with falsehood of political groups reach on average more people at every depth as the cascades were deepened.


\subsection{Temporal Characteristics}
 
We now turn our focus to the length of time of the attention cascades. The first column of Figure~\ref{fig:cascades_static_attributes_part5} displays the CCDF of the cascade duration. The top graph of the this column shows that political cascades last longer than non-political ones.  A possible explanation is that political cascades stir more debate among the participants of the group. The graph at the bottom  of the first column of
Figure~\ref{fig:cascades_static_attributes_part5} shows the CCDF of the duration of  unclassified cascades and cascades with false information. As we can see from the graph, in both types of groups, political and non-political, the presence of false information reduces the duration of the cascades. We conjecture that when false information or fake news is identified in the cascade, the participants of the group start losing interest in the discussion and terminate the cascade.

Regarding temporal cascade attributes as a function of depth and unique users, we observe on the topmost left graph of Figure ~\ref{fig:cascades_dynamic_attributes_time} that cascades in political groups take two times longer than cascades in non-political groups to reach their maximum depth. The topmost right graph indicates that cascades in political groups take considerable more time to reach 90\% of unique users than non-political groups.  Moreover, both graphs are that attention allocation in a WhatsApp group clearly depends on the nature of the content that starts the cascade. One possible explanation is that political discussions spur hot debates that last longer.  In the lower part of Figure~\ref{fig:cascades_dynamic_attributes_time}, we also note that political cascades with falsehood take longer to reach the maximum depth.  However, non-political cascades with falsehood take less time to reach the maximum depth. Although we have not explored the content in messages, we conjecture that political fake news has a significant impact on group discussion and generate more debate than non-political fake news, as shown in the leftmost graph of Figure~\ref{fig:cascades_dynamic_attributes_time}.

\subsection{User Participation in Cascades}

In this section, we focus on the following question:  how participants of a cascade interact with the other participants? We analyze the relative frequency of the six motifs that represent the most common patterns of social communication among participants of the cascades.
Figure~\ref{fig:user_motifs} shows that the most common pattern of communication among participants of a cascade is  \emph{chain}, for both political and non-political groups. \emph{Chain} is the most popular motif in the two classes of groups. The percentage of \emph{chain} is more than half of the total motifs identified in the cascades. It is interesting to observe that the fraction of \emph{self-loops} in political groups is two times greater than the percentage of \emph{self-loops} in non-political groups. Regarding misinformation analysis, the rightmost graph of Figure~\ref{fig:user_motifs} displays that cascades with falsehood of political groups have a higher percentage of chain and self-loops than corresponding non-political cascades. One possible explanation for the higher percentage of self-loops is that in political discussions a participant might want to emphasize or defend her position or idea and makes explicit reference to her previous message in the cascade. 
 
Regarding the total number of users participating in cascades, we note on the second column of Figure~\ref{fig:cascades_static_attributes_part5} that cascades in non-political groups reach more users than in political groups. This characteristic indicates that discussions in non-political groups draw more attention than the ones on political groups probably because of their content. Moreover, cascades with falsehood on political groups reach more users than cascades with falsehood on non-political groups.

\section{Conclusion}
In this paper we used an extensive set of messages to characterize attention cascades in WhatsApp groups from three distinct perspectives: structural, temporal and interaction patterns. We developed a data collection methodology and model each group as a directed acyclic graph, where each connected component is identified as a cascade. In addition to providing a new way of looking at information dissemination in small groups through the concept of attention, our study has unveiled several interesting findings, regarding political groups and misinformation. These findings include differences in the structural and temporal characteristics of political groups when compared to non-political groups. We show that cascades with false information in political groups are deeper, wider and reach more users than cascades with falsehood in non-political groups. We demonstrate that political cascades last longer than non-political ones.  We also show that attention cascades in WhatsApp groups reflect real-world events that capture public attention.

Our current and future work is focused on leveraging many of the findings and conclusions presented in this paper along with several dimensions. First, we are looking into using information about content to understand their impact on the structure and temporal characteristics of attention cascades.  Questions we want to answer in the future include the following: What kind of content would extend the duration of attention cascades? What is an adequate metric for quantifying attention in a group chat?  What is the impact of hate content on the characteristics of
attention cascades? Finally, we want to construct tools to generate synthetic attention cascades to model different types of group chats.

\section{Acknowledgements}
This work was partially supported by CNPq, CAPES, FAPEMIG and the projects InWeb, MASWEB and INCT-Cyber.

\bibliographystyle{ACM-Reference-Format}
\balance
\bibliography{bibliography}

\end{document}